\documentclass[10pt,twocolumn,preprintnumbers,superscriptaddress,nofootinbib,aps,prd]{revtex4-2}

\usepackage{graphicx}
\usepackage{amsmath}
\usepackage{srcltx}
\usepackage[utf8]{inputenc}
\usepackage[colorlinks]{hyperref}
\usepackage{url}
\usepackage{times}
\usepackage{amssymb}
\newcommand{\vev}[1]{\langle {#1} \rangle}
\newcommand{\lsim}{\lesssim}
\newcommand{\gsim}{\gtrsim}

\newcommand{\eq}[1]{Eq.~(\ref{#1})}

\newcommand{\ord}[1]{\mathcal{O}{(#1)}}
\newcommand{\beq}{\begin{equation}}
\newcommand{\eeq}{\end{equation}}
\newcommand{\bea}{\begin{eqnarray}}
\newcommand{\eea}{\end{eqnarray}}
\newcommand{\eps}{\varepsilon}

\newcommand{\mP}{M_{\rm P}}
\newcommand{\rmP}{\bar{M}_{\rm P}}

\newcommand{\appropto}{\mathrel{\vcenter{
  \offinterlineskip\halign{\hfil$##$\cr
    \propto\cr\noalign{\kern2pt}\sim\cr\noalign{\kern-2pt}}}}}

\begin{document}

\pagestyle{plain}

\title{Gravitationally Misaligned Ultralight Dark Matter and Implications for Neutron Stars}

\author{Hooman Davoudiasl}
\email{hooman@bnl.gov} 
\affiliation{High Energy Theory Group, Physics Department \\ Brookhaven National Laboratory,
Upton, NY 11973, USA}


\begin{abstract}
	
We examine the possibility that dark matter (DM) may be an ultralight scalar that was misaligned via non-minimal coupling to gravity, in the early Universe.  For a certain regime of scalar masses, gravitational effects in neutron stars could place interesting bounds on the viable parameter space of the model, even in the absence of non-gravitational interactions between DM and ordinary matter.

\end{abstract}
\maketitle


The dominant material in the Universe, generally referred to as dark matter (DM), is of unknown origin and character.  The evidence for DM is strong -- perhaps irrefutable -- but entirely limited to its gravitational imprint on the visible world \cite{ParticleDataGroup:2022pth}.  As such, many proposals for what it is or how it came about are currently viable, from a phenomenological point of view.  Large classes of ideas require some level of non-gravitational interaction between DM and particles of the Standard Model (SM).  However, there are a number of possibilities that do not require invoking interactions between DM and SM beyond that of gravity.  A well-known example is the case of primordial black holes, produced in the early Universe \cite{Hawking:1971ei,Carr:1974nx,Carr:1975qj,Khlopov:2008qy,Carr:2020xqk}, that are sufficiently long-lived.  

Another class of models employs a light scalar that starts out with some initial displacement, or {\it misalignment}, in its potential and starts to oscillate once its mass exceeds the Hubble expansion rate.  The Peccei-Quinn mechanism -- proposed to address the extreme suppression of CP violation in strong interactions \cite{Peccei:1977hh,Peccei:1977ur} -- leads to the emergence of such a scalar, the axion \cite{Weinberg:1977ma,Wilczek:1977pj}, which is a possible DM candidate \cite{Preskill:1982cy,Abbott:1982af,Dine:1982ah}.  In this case, the axion has feeble couplings to hadrons of the SM.  However, variants of this general idea have been proposed over the years, where no such couplings are required (see, {\it e.g.}, Refs.~\cite{Hu:2000ke,Hui:2016ltb}).  In any event, one needs to assume that some level of initial misalignment existed that set the abundance of DM to its currently observed value. 

In what follows, we will consider an ultralight scalar $\phi$ whose misalignment is generated gravitationally.  This mechanism was introduced for heavy DM production at high mass scales $\gsim \ord{10^6~{\rm GeV}}$ in Ref.~\cite{Babichev:2020xeg}; see also Ref.~\cite{Laulumaa:2020pqi} for an implementation of the idea for MeV scale masses.\footnote{A version of this idea was also employed in Ref.~\cite{Croon:2020ntf} as a means of scanning for resonant production of DM.}   We further assume that the couplings of this scalar to other fields are only of gravitational strength.  We will consider gluon condensation as a possible avenue for generating the tiny mass of $\phi$, denoted by $m_\phi$.  In this case, $\phi$ has no bare mass well above the scale of quantum chromodynamics (QCD) confinement, which we take to be  $\Lambda_{\rm QCD}\sim 200$~MeV, as a rough estimate.  Strictly speaking, though, this is not necessary and one could set $m_\phi$ to the chosen value as a bare parameter.  As we will demonstrate, near the typical values considered here for $m_\phi$, regardless of its origin, the dense interior of a neutron star (NS) may provide a venue to probe the gravitationally misaligned DM postulated in our work.\footnote{Ref.~\cite{Sankharva:2021spi} examines cosmology of non-minimally coupled ultralight axions, with masses $\lsim 10^{-19}$~eV, many orders of magnitude lighter than the scalars considered in our work.  Also, the scalar dynamics we consider has important effects from quartic self-interactions, not included for the axion cosmology of Ref.~\cite{Sankharva:2021spi}; see also Ref.~\cite{Takahashi:2015waa}.  Non-minimal coupling effects on solitons of ultralight DM have been considered in Refs.~\cite{Zhang:2024bjo,Chen:2024pyr}.} 

Here, we would like to add a note on some relevant background work, regarding formation of scalar condensates in NSs which we will consider later.  This possibility is often referred to as {\it scalarization}, first introduced in Ref.~\cite{Damour:1993hw} and later examined \cite{Damour:1996ke} in relation to its effect on binary pulsar measurements.  An extension of this idea was considered in Ref.~\cite{Chen:2015zmx}, where a massive scalar field with generalized coupling was assumed and proposed as a possible dark matter candidate.  Our work shares some elements with this reference, but our underlying scalar model is different.  The non-minimally coupled scalar we have assumed here leads to couplings with ordinary matter akin to those in scalarization models.  The prospects for scalarization to address an apparent puzzle posed by observations of $\approx 2 M_\odot$ NSs, where $M_\odot \approx 2 \times 10^{33}$~g is the solar mass, was investigated in Ref.~\cite{Morisaki:2017nit}\footnote{We thank Teruaki Suyama for pointing out relevant references and comments on the possibility of resolving the 2$M_\odot$ NS puzzle using scalar physics similar to that investigated here.}; see also Ref.~\cite{Sotani:2017pfj}.

To realize our DM model, let us consider the following action for $\phi$
\beq
{\cal I}_\phi =  \int  d^4 x \, \frac{\sqrt{-g}}{2}
\left[\partial^\mu \phi \partial_\mu \phi 
- \left(m_\phi^2 + \xi {\cal R}\right) \phi^2  - \frac{\lambda}{2} \phi^4\right],
\label{Iphi}
\eeq
where $g$ is the determinant of the metric $g_{\mu\nu}$, the mass of $\phi$ is denoted by $m_\phi$, and ${\cal R}$ is the Ricci scalar of general relativity; $\xi > 0$ is a constant and $\lambda > 0$ denotes the quartic coupling of $\phi$.  Our metric signature is taken to be $(+,-,-,-)$.  In what follows, we assume that $m_\phi^2 > 0$.  With these conventions, the Einstein-Hilbert action is given by
\beq
{\cal I}_{\cal EH} = -\frac{1}{16 \pi G_N}\int d^4 x \, \sqrt{-g}\, {\cal R}\,,
\label{IEH}
\eeq 
where we have omitted a possible cosmological constant term.  Here, $G_N= \mP^{-2}$ is Newton's constant and $\mP\approx 1.22 \times 10^{19}$~GeV is the Planck mass.

By Einstein's equations, ${\cal R}$ is proportional to $T^\mu_\mu$, the trace of the energy momentum tensor, according to 
\beq
{\cal R} = - 8 \pi G_N T^\mu_\mu\,.
\label{R-T}
\eeq
For a perfect cosmic fluid, we have $T^\mu_\mu = \rho-3p$, where $\rho$ is the energy density and $p$ is pressure.  Using the equation of state $w \equiv p/\rho$, \eq{R-T} then gives
\beq
{\cal R} = - 3 (1-3w) H^2\,,
\label{R-H2}
\eeq 
where the Hubble expansion rate $H$ is given by $H^2 = (8\pi G_N/3) \rho$ (for a flat Universe). 

We note that $T^\mu_\mu$ is traceless for pure radiation, corresponding to $w=1/3$.  Nonetheless, since the interactions of the SM are not conformal, we expect some deviation from this limit, which we quantify by $\eps>0$ and set 
\beq
{\cal R} = -\eps H^2.  
\label{R}
\eeq
During the radiation dominated era, we have $H = z \,T^2$, where 
\beq
z \equiv \sqrt{\frac{8 \pi^3 g_*}{90 \, \mP^2}}, 
\label{z}
\eeq
with $g_*$ counting the relativistic degrees of freedom and $T$ the temperature.  We may expect $\eps \sim \text{few}\times (0.01-0.1)$ \cite{Davoudiasl:2004gf,Caldwell:2013mox,Croon:2020ntf}.  Hereafter, we will consider cosmological evolution during radiation domination, unless otherwise specified.  

Let us define $\mu_\phi^2 \equiv \xi\, {\cal R} + m_\phi^2$.  From \eq{Iphi}, we may deduce that for $\mu_\phi^2 < 0$ the field $\phi$ develops a vacuum expectation value (vev)
\beq
\vev{\phi} = \sqrt{\frac{|\mu_\phi^2|}{\lambda}}\,,
\label{vevphi}
\eeq
which decreases with falling temperature until $\mu_\phi^2\geq 0$, and thereafter $\vev{\phi}=0$.  To have $\phi$ track the above vev, its time evolution cannot be too rapid.  This corresponds to the requirement \cite{Babichev:2020xeg}
\beq
\frac{1}{\mu_\phi^2}\frac{d \mu_\phi}{dt} \sim \frac{H}{\mu_\phi} \ll 1\,.
\label{slow-evol}
\eeq
The above would then be satisfied if $\mu_\phi \gg H$.  For $|\xi {\cal R}| \gsim m_\phi^2$ ({\it i.e.} $\vev{\phi}\neq 0$), this implies 
\beq
\sqrt{\xi\, \eps} \gg 1 \quad ; \quad \text{(vev tracking condition)}.  
\label{xi-eps}
\eeq
We note that the requirement (\ref{slow-evol}) for slow evolution of $\vev{\phi}$ would cease to hold as $|\xi {\cal R}| \to  m_\phi^2$.  After this point, $\vev{\phi} = 0$, around which $\phi$ would oscillate once Hubble friction is not efficient, corresponding to $m_\phi \gsim 3 H$.

We will later address possible constraints on the assumption (\ref{xi-eps}) and the regime of validity of the scenario considered here.  However, a few comments on the above requirement implying $\xi\gg 1$ are warranted.  Working in a regime where $\phi$ tracks its vev leads to a thermal evolution that is not governed by the kinetic energy of $\phi$ before it starts to oscillate and approach DM behavior.  This makes it feasible for us to derive simple analytic approximations for the requisite parameters in a viable DM scenario.  One could, in principle, choose $\xi \sim 1$ and study the resulting dynamics.  However, we found that for such values of $\xi$ it is not straightforward to infer the approximate required parameters, prior to obtaining numerical solutions.  Nonetheless, in this regime of $\xi$, one still generally expects to find values of parameters where reasonable DM solutions obtain, leading to different implications for the model and its domain of validity.  We choose to examine the dynamics  which requires $\xi \gg 1$, where a more transparent treatment and clearer physical picture seem to result.\footnote{Early work where large values of $\xi$ were considered in different contexts can be found, for example, in  Refs.~\cite{Spokoiny:1984bd,Fakir:1990eg}; see also Ref.~\cite{Bezrukov:2007ep} where $\xi \gg 1$ was considered in a model of Higgs inflation.}     

Let us denote the temperature at which $\vev{\phi}\to 0$ by $T_*$.  At $T=T_*$, we have 
\beq
\xi \eps H_*^2 \approx m_\phi^2\,, 
\label{H*}
\eeq
where $H_* \equiv H(T_*)$. We thus obtain 
\beq
T_* \approx \left(\frac{m_\phi}{z \sqrt{\xi\,\eps}}\right)^{1/2}.  
\label{T*}
\eeq
Since we are assuming that $\sqrt{\xi \,\eps} \gg 1$, the field $\phi$ would start its oscillation at $T\sim T_*$, as long as $\sqrt{\xi \eps}\gsim 3$, which implies $m_\phi \gsim 3 H_*$, by \eq{H*}.  Following the discussion in Ref.~\cite{Babichev:2020xeg}, one can show that the initial misalignment of $\phi$ is given by 
\beq
\phi_* \approx \left(\frac{2}{\xi \, \eps}\right)^{1/6} \frac{m_\phi}{\sqrt{\lambda}}\,.
\label{phi*}
\eeq
For $(\xi \, \eps)^{1/3} \gsim 1$, we see that $\lambda \phi_*^2 \lsim m_\phi^2$, however at $T=T_*$ the mass parameter $\mu_\phi^2$ vanishes.  Nonetheless, slightly below $T_*$, we expect the contribution from ${\cal R}$ which scales with $T^4$, to become sufficiently small.  Hence, for $T\lsim T_*$ the evolution of $\phi$ is largely governed by its bare mass term $m_\phi^2 \phi^2/2$, as will be illustrated below, and the scalar energy density enters matter-like regime as required for DM.  We also note that any dominant kinetic contribution to $\phi$ energy density near $T=T_*$ will redshift quickly.  For the sake of concreteness, we will set 
\beq
\eps=0.1 \;\; \text{and} \;\; \xi = 10^3\,, 
\label{eps+xi}
\eeq
in what follows. 

Let us now estimate the required parameters for $\phi$ to be a viable DM candidate.  We will take the energy density at the time of matter-radiation equality to be $\rho_{\rm eq} \approx$~eV$^4$ at $T_{\rm eq}\approx$~eV.  This is an approximation, but since we only intend to show the viability of the model up to $\ord{1}$ effects, it will suffice for our purposes.  We then demand 
\beq
\rho_{\rm eq} \approx \left(\frac{T_{\rm eq}}{T_*}\right)^3\rho_*\,,
\label{rhoeq}
\eeq
where  
\beq
\rho_*\approx \frac{1}{2} m_\phi^2 \phi_*^2\,.
\label{rho*}
\eeq
Using Eqs.~(\ref{rhoeq}) and (\ref{rho*}), we obtain 
\beq
\lambda \approx  \frac{z^{3/2} \,(\xi \, \eps)^{5/12}\, m_\phi^{5/2}}{4^{1/3}\,T_{\rm eq}} 
\quad ; \quad (\text{$\phi$ as DM}).
\label{lambda}
\eeq
The above equation estimates the required value of $\lambda$ for viable DM, given input values for $g_*$, $\xi$, $\eps$, and $m_\phi$.  

{\it Regime of validity:} Before going further, let us outline the regime of validity of our treatment.  We will assume that the minimally coupled SM (and perhaps other) fields are the main components of $T^\mu_\mu$ and that the non-minimal coupling of $\phi$ does not lead to major deviations from this assumption.  Given the large value of $\xi$ adopted above, one needs to determine the range of parameters for which this condition holds.  Using the results of Ref.~\cite{Figueroa:2021iwm}, the contribution to $T^\mu_\mu$ from the non-minimally coupled $\phi$ is given by
\beq
T^\mu_{\mu(\phi)} = (6\xi -1)\left[\partial_\mu \phi \partial^\mu \phi + \mu_\phi^2 \phi^2 + \lambda \phi^4\right]-m_\phi^2 \phi^2\,,
\label{Tracephi}
\eeq 
which vanishes in the conformal limit $\xi \to 1/6$ and $m_\phi\to 0$.  However, as we have $\xi\gg 1$, we need to make sure that the above does not overwhelm the contribution $T^\mu_{\mu(m)}$ of minimally coupled fields \footnote{We thank A. Florio for emphasizing this point to us.}.

Before the onset of oscillation, for $T\gsim T_*$, we may assume $\phi \approx \sqrt{|\xi {\cal R}|/\lambda}$ and hence 
\beq
\frac{d\phi}{dT}\approx 2 z \sqrt{\frac{\xi \eps}{\lambda}}\, T\,,
\label{dphidT}
\eeq
using \eq{R}.  During radiation domination, we have $dT/dt = - H^2/(z T)$ and therefore 
\beq
\dot{\phi}^2 \approx 4 H^2 \phi^2.
\label{phidot2}
\eeq 
In the limit of $\vev{\phi}\neq 0$ tracking, the second and third terms in the square bracket in \eq{Tracephi} are of equal magnitude and cancel out.  Thus, the kinetic contribution to $T^\mu_{\mu(\phi)}$ is dominant and given by $6\xi \dot{\phi}^2$, where we have assumed $\xi\gg 1$ and ignored spatial gradients in the primordial background.  Demanding that minimally coupled matter dominates, {\it i.e.} $6\xi \dot{\phi}^2 \ll T^\mu_{\mu(m)}$, then yields the condition
\beq
\phi \ll \sqrt{\frac{\eps}{24 \xi}}\, \rmP\,.
\label{val-cond}
\eeq
With our benchmark values set above, we have $\phi \ll 5\times 10^{15}$~GeV, as the regime of validity for our approximations, which corresponds to treating the non-minimal coupling of $\phi$ as a perturbation.

To go further, we will fix some of the parameters.  Let us begin with $m_\phi$, which we will assume to be in the ultralight regime, $\ll$~eV, but otherwise free.  However, to provide a benchmark, we will consider the possibility that it is set by QCD confinement via gravitational effects, characterized by the effective dimension-6 interaction
\beq
O_6 = c_6 \frac{\phi^2}{\mP^2} G_{\mu\nu}^a G^{a\mu\nu}.
\label{O6}
\eeq
In \eq{O6}, $c_6$ is a constant, $G_{\mu\nu}^a$ is the gluon field strength tensor, and $a=1,2,\ldots,8$ is the adjoint index.  Upon confinement, the gluon condensate takes a value \cite{Shifman:1978bx,Shifman:1978by,Gubler:2018ctz} $\vev{\alpha_s G_{\mu\nu}^a G^{a\mu\nu}}\sim 0.03-0.3$~GeV$^4$, where $\alpha_s$ is the QCD fine structure constant.  The condensate value is not very precisely determined \cite{Gubler:2018ctz}, but we are only using it as a rough estimator: 
\beq
m_\phi^2 \sim c_6 \frac{\vev{G_{\mu\nu}^a G^{a\mu\nu}}}{\mP^2}.
\label{mphi}
\eeq
Setting $c_6=1$, we then adopt $m_\phi = 10^{-11}$~eV as a benchmark, motivated by the above considerations.  We will see that for $m_\phi$ near this scale interesting astrophysical probes may arise, though the origin of this mass scale is not directly relevant to that phenomenology.

Here, we also briefly mention that if there is a Planck-suppressed dimension-8 operator contributing to the quartic coupling of $\phi$, we expect $\delta \lambda \sim \vev{\alpha_s G_{\mu\nu}^a G^{a\mu\nu}}/\mP^4\sim 10^{-78}$.  One can check that, for the range of $m_\phi$ allowed by the bound (\ref{mphi-NS-const}), this correction would be negligible compared to the values implied by \eq{lambda}.

Based on the above assumptions, the oscillations of $\phi$ only begin after QCD confinement, which implies $T_* \lsim 200$~MeV.  Hence, we will take $g_* = 10.75$, as a representative value; \eq{T*} then gives $T_*\approx 50$~MeV.   From \eq{lambda}, we find $\lambda \approx 1.3 \times 10^{-68}$, which yields $\phi_* \approx 4.6 \times 10^{13}$~GeV; this is consistent with the requirement (\ref{val-cond}). 

Adapting the results in Ref.~\cite{Croon:2022gwq}, one can straightforwardly show that \eq{Iphi} yields the following equation for the thermal evolution of the $\phi$ background value during the radiation dominated era  
\beq
z^2\, T^6 \frac{d^2\phi}{dT^2} + 
\left(m_\phi^2 - \xi\, \eps\, z^2\, T^4 \right) \phi + \lambda\, \phi^3 = 0\,.
\label{phi-Tevol}
\eeq
It is implied here that the relation (\ref{val-cond}) is satisfied.  
We note that $g_*$ varies with temperature, as the Universe cools below various particle thresholds.  However, in our analysis we will choose $g_* = 10.75$, for simplicity, based on the preceding discussion.  Given the possible range of parameters, it suffices for the purposes of this work to show that the model can provide a fair approximation to the observed properties of DM, up to $\lsim \ord{1}$ corrections, as mentioned earlier.

\begin{figure}[t]\vskip0.25cm
	\centering
	\includegraphics[width=\columnwidth]{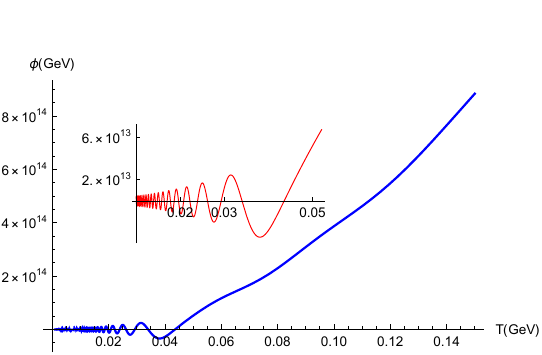}
	\caption{Evolution of $\phi$ background with temperature, in GeV units.  The inset plot shows the same solution, but for a limited interval above $T=0.01$~GeV.  Here, we have set $\xi=10^3$, $\eps=0.1$, and $m_\phi= 10^{-11}$~eV.  The value of $\lambda$ is given by \eq{lambda}.  The initial temperature is assumed to be 0.15 GeV, for the presented solution.}
	\label{phi-evol}
\end{figure}

\begin{figure}[t]\vskip0.25cm
	\centering
	\includegraphics[width=\columnwidth]{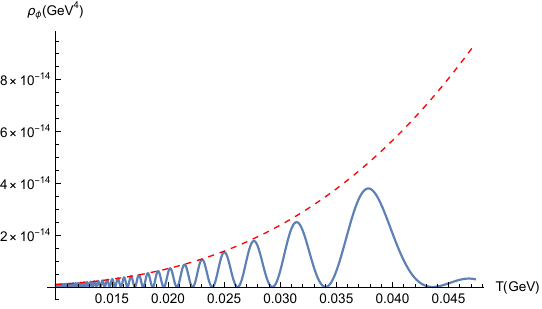}
	\caption{Evolution of DM energy density $\rho_\phi= \mu_\phi^2 \phi^2/2 + \lambda \phi^4/4$ (solid curve) with temperature.  The dashed curve corresponds to scaling with $T^3$, which is the correct behavior for matter-like energy density.  Same model parameters as in Fig.~\ref{phi-evol} have been taken.}
	\label{rhophi}
\end{figure}

In Fig.~\ref{phi-evol}, we show the behavior of the $\phi$ background, corresponding to gravitationally induced $\vev{\phi}$, for the above benchmark choices.  The initial values chosen correspond to $\vev{\phi}$ in \eq{vevphi} and its derivative with respect to $T$ at $0.15$~GeV.   The estimates from \eq{T*} and \eq{phi*} suggest that the oscillations would start at $T_* \sim 50$~MeV with an initial misalignment of $|\phi_*|\sim 4.6\times 10^{13}$~GeV, which is approximated by the curve shown in the inset, reasonably well.  

We show the behavior of the energy density, defined here as $\rho_\phi = \mu_\phi^2 \phi^2/2 + \lambda \phi^4/4$ above and near $T=0.01$~GeV in Fig.~\ref{rhophi}.  One expects that $\rho_\phi$ at $T=0.01$~GeV, well below $T_*$, to be $\sim 10^{-15}$~GeV$^4$, based on $T^3$ scaling; this expectation is consistent with the value shown.  Also, we see that $\rho_\phi$ exhibits the expected $T^3$ scaling, corresponding to the dashed line envelope in Fig.~\ref{rhophi}, fairly quickly as $T$ falls below $T_*$,  in agreement with our earlier qualitative analysis.

At this point, we would like to examine what possible observational bounds could be placed on the DM scenario presented in this work.  Our only possible coupling to the SM so far has been through a Planck-suppressed interaction, potentially induced by non-perturbative gravity, that we used as a heuristic guide for the mass scale of our ultralight DM candidate.  However, this is not a necessary assumption and we may simply set $m_\phi$ at any value as a model input parameter.  In that case, it would seem that constraining the model would not be feasible.  However, as we will discuss below, the dense environment of a neutron star (NS) could provide a possible probe of the DM candidate proposed in this work.

We will consider a ``typical" NS of mass $1.4 M_\odot$.  Across most of the NS, it is estimated that the Ricci scalar assumes a value ${\cal R}_{\rm NS}\approx - 10^{-12}$~cm$^{-2}$ \cite{Eksi:2014wia}\footnote{Note that our metric signature convention is different from that in Ref.~\cite{Eksi:2014wia}, hence the opposite sign for the Ricci scalar.}.  With our benchmark value $\xi = 10^3$, we have $\sqrt{|\xi\, {\cal R}_{\rm NS}|} \approx 6 \times 10^{-10}$~eV.  For $m_\phi < \sqrt{|\xi {\cal R}_{\rm NS}|}$, we then expect that $\vev{\phi}\neq 0$ inside the NS 
\beq
\vev{\phi}_{\rm NS} \approx \sqrt{\frac{|\xi\, {\cal R}_{\rm NS}+m_\phi^2|}{\lambda}}.
\label{phiNS}
\eeq
We note that by Einstein's equations ${\cal R} = 0$ in vacuum, corresponding to the exterior of the NS, and hence $\vev{\phi}=0$ outside the star, to a very good approximation.  We may expect that if $\vev{\phi}_{\rm NS}$ gets sufficiently large, it would cause the NS to develop non-standard features.  

A detailed analysis of the evolution of the NS after a core collapse supernova in the presence of the new physics postulated here is outside the scope of our work.  Also, the precise properties of the hadronic medium inside an NS are not well-understood and are subject to significant uncertainty \cite{Lattimer:2021emm}, though progress is ongoing (see, {\it e.g.}, Ref.~\cite{Rutherford:2024srk}).  We thus simply demand that the apparent contribution of $\phi$ inside the NS not be large enough to become a significant perturbation to standard physics.  

To estimate the new physics contribution, we assume that $\vev{\phi}=0$ before the collapse of the progenitor star.  Once the stellar density, or equivalently ${\cal R}$, is large enough to cause a transition to $\vev{\phi} = \phi_{\rm NS}$, the field will roll to its new minimum, as in a Higgs mechanism, and start to oscillate. Since all the couplings of $\phi$ are tiny, these oscillations will persist without significant dissipation, converting potential energy of $\phi$ to kinetic energy $\dot{\phi}^2/2$, and vice versa.  The potential energy difference is 
\beq
|V(0) - V(\phi_{\rm NS})| = \frac{\lambda}{4} \phi_{\rm NS}^4, 
\label{potential-diff}
\eeq
and hence, the maximum kinetic energy is given by $\dot{\phi}^2/2 = \lambda \phi_{\rm NS}^4/4$.  Near the minimum of the potential, the stress-energy tensor is dominated by the kinetic contribution.  Using \eq{Tracephi}, we then have     
\beq
T^\mu_{\mu(\phi)}\approx 6\xi \dot{\phi}^2 \approx 3 \xi \lambda \phi_{\rm NS}^4.   
\label{Tracephi-NS}
\eeq

Taking ${\cal R}_{\rm NS}=T^\mu_{\mu({\rm NS})}/\rmP^2 \approx (1.97 \times 10^{-20}~\text{GeV})^2$ \cite{Eksi:2014wia}, we obtain $T^\mu_{\mu({\rm NS})} \approx 2.30 \times 10^{-3}$~GeV$^4$ for the trace of $T_{\mu\nu}$ inside the NS.   Let us define 
\beq
r_\phi \equiv \frac{T^\mu_{\mu(\phi)}}{T^\mu_{\mu({\rm NS})}}\,,
\label{rphi}
\eeq
as the ratio of a typical $\phi$ contribution to the energy budget of the NS compared to that of standard physics, which implies $r_\phi \approx 3 \xi \lambda \phi_{\rm NS}^4/T^\mu_{\mu({\rm NS})}$.  Demanding $r_\phi < 1$, we find

\beq
m_\phi \gsim 3.0 \times 10^{-11}~\text{eV} \quad \text{(NS Constraint)}. 
\label{mphi-NS-const}
\eeq      

The above lower bound is quite close to the value of $m_\phi$ adopted before, assuming the Planck-suppressed gluon confinement contribution, $m_\phi \sim 10^{-11}$~eV, in \eq{O6}.  Hence, we may assume that the suggested QCD-induced mechanism, up to mild modulations of its strength, can be the source of $m_\phi$ without leading to large deviations from standard NS physics.  Also, the above suggests that, for our benchmark input parameters, NS properties may receive non-negligible contributions from the dynamical response of $\phi$ to gravity.  

\begin{figure}[t]\vskip0.25cm	
	\includegraphics[width=\columnwidth]{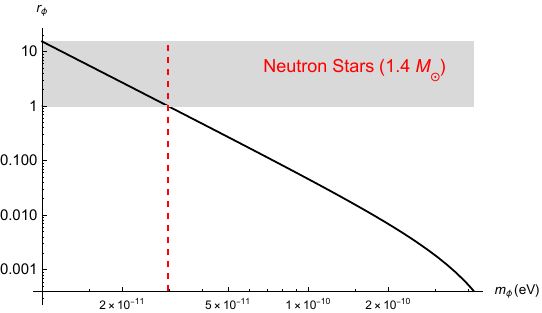}
	\caption{The ratio $r_\phi \equiv 3 \xi \lambda \phi_{\rm NS}^4/T^\mu_{\mu({\rm NS})}$ in \eq{rphi}.   The shaded gray region at the top of the plot corresponds to $r_\phi \geq 1$ that we exclude to avoid large departures form the expected properties of neutron stars, assuming a mass of $1.4 M_\odot$.  Here, the parameter space corresponds to $\phi$ being DM.  The red dashed vertical line marks the minimum value of $m_\phi$ allowed by this requirement.  The region of parameters to the left of this line is thus disfavored in our DM model.  As before, we have assumed $\xi=10^3$ and $\eps =0.1$.  See the main text for more details.}
	\label{NS-Bound}
\end{figure}

We have plotted the ratio $r_\phi$ in Fig.~\ref{NS-Bound} as a function of $m_\phi$, assuming that it could be viable DM.  The shaded gray area in this figure corresponds to parameters for which the contribution of $\phi$ to the NS internal dynamics becomes equal or larger than that of ordinary matter.  Likely, this contribution needs to be much smaller than $\ord{1}$, and we can take the obtained constraint as quite conservative.

Let us also note that for other choices of NS parameters, one may find {\it qualitatively different} implications.  For example, it is possible that the trace anomaly of the hadronic matter inside the NS, for sufficiently high energy densities, could be negative \cite{Fujimoto:2022ohj}.  This would switch off $\vev{\phi}$ and remove the induced correction to $T^\mu_{\mu({\rm NS})}$, discussed earlier.  In that case, one may have a circumstance in which the effect considered here would be manifested only in some population of NSs or possibly over a certain region inside the star.  This could lead to an interesting handle on our scenario, which would warrant further investigation in future work.

\begin{figure}[t]\vskip0.25cm
	\centering
	\includegraphics[width=\columnwidth]{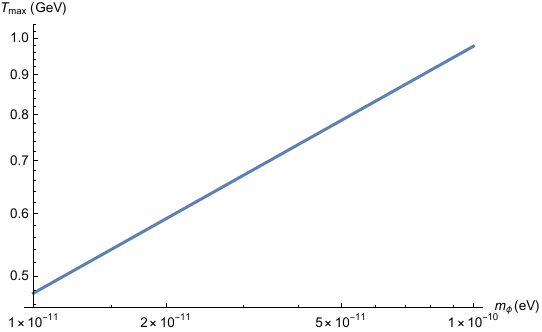}
	\caption{The maximum temperature for which the value of $\vev{\phi}$ induced by ${\cal R}$ in the early Universe does not lead to saturation of the inequality (\ref{val-cond}).}
	\label{Tmaximum}
\end{figure}

Another possible astrophysical handle on ultralight DM in the mass range examined here is based on observations of black holes (BHs) with significant spin.  If there is a boson whose Compton wavelength is comparable to the Schwarzschild radius of the BH, a cloud of that boson would be populated around it by depleting its rotational energy \cite{Penrose:1971uk,1971JETPL..14..180Z,Misner:1972kx,Starobinsky:1973aij}.  This process is often referred to as {\it superradiance}; see, {\it e.g.}, Ref.~\cite{Brito:2015oca} for a review of this subject.  It was suggested in Ref.~\cite{Arvanitaki:2009fg} that superradiance could provide a probe of ultralight bosons. Self-interactions of  bosons can, however, reduce the efficiency of superradiance.  This effect was studied in detail for the case of scalars in Ref.~\cite{Baryakhtar:2020gao}, which updated the earlier results in Ref.~\cite{Arvanitaki:2014wva}.  The analysis of Ref.~\cite{Baryakhtar:2020gao} roughly excludes scalars of mass $2.5 \times 10^{-13}~\text{eV}\lsim m_\phi \lsim 3.5 \times 10^{-12}~\text{eV}$.  However, this range is already disfavored in our model by the above NS bound.  

Let us now come back to the question of the maximum temperature for which our analysis can be valid.  This corresponds to avoiding values of $\vev{\phi}$ which can perturb standard physics by $\ord{1}$ effects at sufficiently high $T$.  In Fig.~\ref{Tmaximum}, we have plotted the maximum temperature 
\beq
T_{\rm max} \approx \left(\sqrt{\frac{\lambda}{24}}\, \frac{\rmP}{\xi z}\right)^{1/2} \,, 
\label{Tmax}
\eeq
corresponding to saturation of requirement (\ref{val-cond}), versus $m_\phi$.  As before, we have assumed parameters consistent with $\phi$ being DM.  Here, we have taken $\phi\sim \sqrt{\xi {\cal R}/\lambda}$ and $g_* = 65$, as a representative value for this temperature regime.  As one can see from the plot, at the lower bound (\ref{mphi-NS-const}) for $m_\phi$ we have $T_{\rm max} \sim 0.7$~GeV, with other maximum reheat temperatures being higher.  If we consider staying a factor 
of $\sim 10$ below saturation of the condition (\ref{mphi-NS-const}), 
we would find $T_{\rm max} \sim 0.2$~GeV.  Given that such values are well above the required minimum reheat temperature of a few MeV \cite{Hannestad:2004px,Ichikawa:2005vw,Hasegawa:2019jsa}, there is no established phenomenological obstruction for having a cosmology consistent with observations in our scenario.

Finally, one may worry that having a non-minimally coupled ultralight scalar could lead to unacceptable violations of equivalence principle through a new long-range force.  However, as we will argue below,  this is not a problem for our model.  To see this, one needs to go from the ``Jordan frame" to the ``Einstein frame" in which the scalar is minimally coupled and gravity is canonical.  This can be done by performing a conformal transformation of the metric
\beq
g_{\mu\nu} \to \Omega(\phi)\, g_{\mu\nu}\,,
\label{conformal}
\eeq    
with 
\beq
\Omega(\phi) \equiv 1 + \xi \frac{\phi^2}{{\bar M}_P^2}.
\label{Omega}
\eeq      
Even without any direct coupling of $\phi$ to the SM, the dependence of the matter action on the metric and its determinant would induce a coupling to the SM content of the form $\xi \phi^2/{\bar M}_P^2$; we will discuss this point in the scholium below.         

The ``loop-level" exchange of $\phi^2$ between SM states would yield an extremely weak potential suppressed by $\xi^2/{\bar M}_P^4$, which is completely negligible.  However, if $\phi$ has a large background value $\phi_0$, one could expand the quadratic term $\phi^2 \to 2 \xi \phi_0 \phi + \ldots$ which yields a linear coupling suppressed by $1/{\bar M}_P$, with a coefficient $2 \xi \phi_0/{\bar M}_P$.  Since $\phi$ is assumed to be DM, its local amplitude oscillates at a frequency set by $m_\phi \sim 10^{-11}~\text{eV} \sim 10^4$~Hz.  Note that $\phi$ does not have a curvature induced expectation value $\vev{\phi}$ in vacuum and the DM background is its only non-trivial source outside astronomical objects\footnote{For $m_\phi$ values considered in this work, the contribution of ${\cal R}$ induced inside  stellar objects less dense than NSs would be too small to lead to $\vev{\phi}\neq 0$.}.  Hence, over macroscopic time scales $\Delta t$ -- associated with equivalence test experiments -- the time average of $\phi_0$ is given by
\beq
\bar \phi_0 = \frac{1}{\Delta t} \int^{\Delta t} dt \, \phi(t) \propto (m_\phi \Delta t)^{-1}\,,
\label{phi0bar}
\eeq
which implies $\bar \phi_0 \to 0$ as $\Delta t\to \infty$, and no net Yukawa potential is obtained.  Note that $A_0 \sim \sqrt{\rho_{\rm DM}}/m_\phi$, where $A_0$ is the amplitude of $\phi_0$ and $\rho_{\rm DM}\sim 0.3-0.4$~GeV cm$^{-3}$ is the DM energy density near the Solar System \cite{Salucci:2010qr,ParticleDataGroup:2022pth}.  This suggests that $A_0\sim 0.1$~GeV and hence, for $\xi\sim 10^3$, any oscillatory force between bodies is suppressed by $(\xi A_0/{\bar M}_P)^2 \sim 10^{-33}$ compared to gravity, rendering it essentially undetectable. 

{\it Scholium:} Here, we provide a schematic illustration of the induced tachyonic mass term for $\phi$,  in the Einstein frame.  Let us consider a minimally coupled free scalar $\Phi$ of mass $m_\Phi$ in the Jordan frame 
\beq
I_J = \frac{1}{2} \int d^4 x \sqrt{-g}\, \left[g^{\mu\nu} \partial_\mu \Phi \partial _\nu \Phi - m_\Phi^2 \Phi^2\right]\,.
\label{Phi-J}
\eeq
The energy momentum tensor, corresponding to a Lagrangian ${\cal L}$, is given by $T_{\mu\nu} = 2\,  \partial {\cal L}/\partial g^{\mu\nu} - g_{\mu\nu} {\cal L}$.  The trace of $T_{\mu\nu}$ for the action in \eq{Phi-J} is then  
\beq
T^\mu_{\mu(\Phi)} = - \partial^\mu \Phi \partial _\mu \Phi + 2 m_\Phi^2 \Phi^2\,.
\label{TracePhi}
\eeq
Going to the Einstein frame, defined by the metric $\hat{g}_{\mu\nu} = \Omega g_{\mu_\nu}$, we have 
$\sqrt{-g} = \Omega^{-2} \sqrt{-\hat{g}}$, and $g^{\mu\nu} = \Omega\, \hat{g}^{\mu\nu}$; here $\Omega \equiv \Omega(\phi)$.  Hence, we obtain the Einstein frame action
\beq
I_E = \frac{1}{2} \int d^4 x \sqrt{-\hat{g}} \, \Omega^{-2} \left[\Omega \,\hat{g}^{\mu\nu} \partial_\mu \Phi \partial _\nu \Phi - m_\Phi^2 \Phi^2\right]\,.
\label{Phi-E}
\eeq
For small $\xi \phi^2/\rmP^2$, we have 
\bea
I_E &\to& I_J|_{g\to \hat g}
\nonumber \\
&+&\frac{1}{2} \int d^4 x \sqrt{-\hat{g}}\, \frac{\xi \phi^2}{\rmP^2}\, [-\partial^\mu \Phi \partial _\mu \Phi + 2 m_\Phi^2 \Phi^2] + \ldots \nonumber \\
&=& I_J|_{g\to \hat g}+ \frac{1}{2} \int d^4 x \sqrt{-\hat{g}}\, \xi  \frac{T^\mu_{\mu(\Phi)}}{\rmP^2} \phi^2+ \ldots.
\label{Phi-E-expand}
\eea
The above yields the Einstein frame negative mass squared parameter for $\phi$, equivalent to that induced by the non-minimal coupling to ${\cal R}$ in the Jordan frame, assuming that the energy momentum tensor is dominated by $\Phi$.  Note that the condition (\ref{val-cond}) implies $\Omega \approx 1$ and hence the form of the $\phi$ potential in the two frames are approximately the same (see, {\it e.g.}, Ref.~\cite{Bezrukov:2007ep}), leading to $\vev{\phi} \neq 0$ as before.  

{\it Summary:} In this work, we proposed that DM may be an ultralight scalar with non-minimal coupling to gravity, through the Ricci scalar of general relativity.  This could allow for gravitational misalignment of the DM during the radiation dominated era and set up the requisite conditions for a viable DM candidate.  While no direct couplings for DM to ordinary matter are required, the model can still lead to phenomenological implications in neutron stars, for certain DM regime of masses.  Such mass scales may be induced by gravitational strength coupling of the scalar to the QCD gluon condensate, though the astrophysical phenomenology considered here does not depend on the physical origin of the scalar mass.  It would be interesting to consider further implications of this DM scenario for the physics of core-collapse supernovae and neutron star formation\footnote{Work along this general direction can be found, for example, in Refs.~\cite{Novak:1999jg,Sperhake:2017itk,Kuroda:2023zbz}.}. 

\vskip0.5cm
{\small \tt All plots presented here can be reproduced from the expressions included in the paper, using standard numerical software.}      

~
\begin{acknowledgments}
We thank Adrien Florio and Kenji Fukushima for helpful discussions of key topics relevant to our work, and Matt Sullivan for related conversations.   We are grateful to Adrien Florio for comments on a draft of the manuscript.  We also thank Teruaki Suyama for communication regarding our work and bringing relevant references to our attention.  This work is supported by the US Department of Energy under Grant Contract DE-SC0012704.
\end{acknowledgments}

\bibliography{R-DM.bib}

\end{document}